\documentclass[12pt]{iopart}

\usepackage{amssymb}
\usepackage{comment}
\usepackage{graphicx}
\usepackage{hyperref}

\begin{document}

\title{Advanced strategies for ion acceleration using high power lasers}

\author{A Macchi$^{1,2}$, A Sgattoni$^{1,3}$, S Sinigardi,$^4$, M Borghesi$^{5,6}$, and M~Passoni$^3$}

\address{$^1$Istituto Nazionale di Ottica, Consiglio Nazionale delle Ricerche (CNR/INO), U.O.S. ``Adriano Gozzini'', Pisa, Italy}
 \ead{andrea.macchi@ino.it}
\address{$^2$Dipartimento di Fisica ``Enrico Fermi'', Universit\`a di Pisa, Largo Bruno Pontecorvo 3, I-56127 Pisa, Italy}
\address{$^3$Dipartimento di Energia, Politecnico di Milano, Via Ponzio 34/3, I-20133 Milan, Italy}
\address{$^4$Dipartimento di Fisica e Astronomia and INFN, Universit\`a di Bologna, via Irnerio 46, 40126 Bologna, Italy} 
\address{$^5$Centre for Plasma Physics, The Queen's University of Belfast, BT71NN Belfast, UK}
\address{$^6$Institute of Physics of the ASCR, ELI-Beamlines Project, Na Slovance 2, 18221 Prague, Czech Republic}

\begin{abstract}
A short overview of laser-plasma acceleration of ions is presented. 
The focus is on some recent experimental results and related theoretical work 
on advanced regimes. These latter include in particular 
target normal sheath acceleration using ultrashort low-energy pulses
and structured targets, 
radiation pressure acceleration in both thick and ultrathin targets, 
and collisionless shock acceleration in moderate density plasmas. 
For each approach, open issues and the need and potential for further 
developments are briefly discussed. 
\end{abstract}


\submitto{\PPCF}

\section{Introduction}

Ion acceleration driven by superintense laser-plasma interaction with dense 
materials has been attracting an enormous 
interest in the last thirteen years because of the several foreseen 
applications such as fast ignition, medical hadrontherapy, nuclear and particle
physics. Thanks to the continuous progress in high power laser technology and 
also in target manufacturing and engineering, several different acceleration 
mechanisms have been either demonstrated or proposed. Still, it is an open 
question to establish which mechanism is most promising for each application, 
as most of the requirements for ion energy, conversion efficiency, spectral 
width, brilliance and suitability for high repetition rate operation have yet 
to be met. 

In this paper we provide a partial update of our recent review of the 
field \cite{macchiRMP13} (see also \cite{daidoRPP12} for a complementary
review) focusing on a selection of most recent results and emerging topics,
and highligthing some contributions by our group.
The emphasis is on basic acceleration mechanisms and general strategies to  
improve their performance.

\section{Advances in Target Normal Sheath Acceleration of protons}

Target Normal Sheath Acceleration (TNSA) 
is recognized as the basic mechanism of acceleration of protons from 
solid targets (typically in the $0.1-10~\mu\mbox{m}$ thickness range)
in most of the experiments reported so far, and has become the reference
framework for source optimization. In TNSA, the protons are accelerated by
the sheath field generated at the rear surface of the target by the 
energetic ``fast'' electrons accelerated at the front surface, where the 
laser-plasma interaction occurs. Thus, the enhancement of the maximum energy 
and conversion efficiency of protons is strictly related to the mechanisms 
of fast electron acceleration and transport through the target.
Similarly, ``engineering'' of TNSA accelerating field by particular target
shaping has been used for proton beam focusing and manipulation.

\subsection{TNSA with ultrashort pulses}

Several groups have investigated proton acceleration using laser systems
with pulse durations of few tens of fs, typically delivering a total
energy or a few Joules. Such laser systems have a compact size and 
potential for high repetition rate as required by most applications. 
The use of very tight focusing, producing laser spots with diameter of the
order of one wavelength $\lambda$, combined with the ultrashort duration 
allows to reach intensities up to $10^{21}~\mbox{W cm}^{-2}$.
The availability of pulses with very high contrast allows using very thin 
targets (from a few $\mu\mbox{m}$ down to tens of~nm) which in principle
may yield higher densities and/or temperatures of fast 
electrons, and thus higher accelerating fields, due to
the effects of energy confinement and electron recirculation. 

Here we analyse a set of experiments with ultrashort pulses in an attempt to
infer scaling and trends of the proton energy versus laser and target 
parameters. 
The data are taken from experiments performed in different laboratories i.e. 
HZDR Dresden (DRACO laser) \cite{zeilNJP10,zeilNC12},
LLC Lund \cite{neelyAPL06}, 
GIST/APRI Gwangju (LiFSA) \cite{choiAPL11,margaronePRL12}, 
JAEA/KPSI Kyoto (JKAREN) \cite{oguraOL12} and CUOS Ann Arbor (HERCULES)
\cite{dollarPoP13}.
This  analysis is meant to update and be complementary to previously
published surveys of data from different experiments and related inferred 
scalings, including comparison with theoretical models 
(see e.g. \cite{daidoRPP12,zeilNJP10,fuchsNP06,borghesiPPCF08,passoniNJP10,peregoNIMA11,kieferPRE13}).

\begin{figure}[t!]
\includegraphics[width=0.99\textwidth]{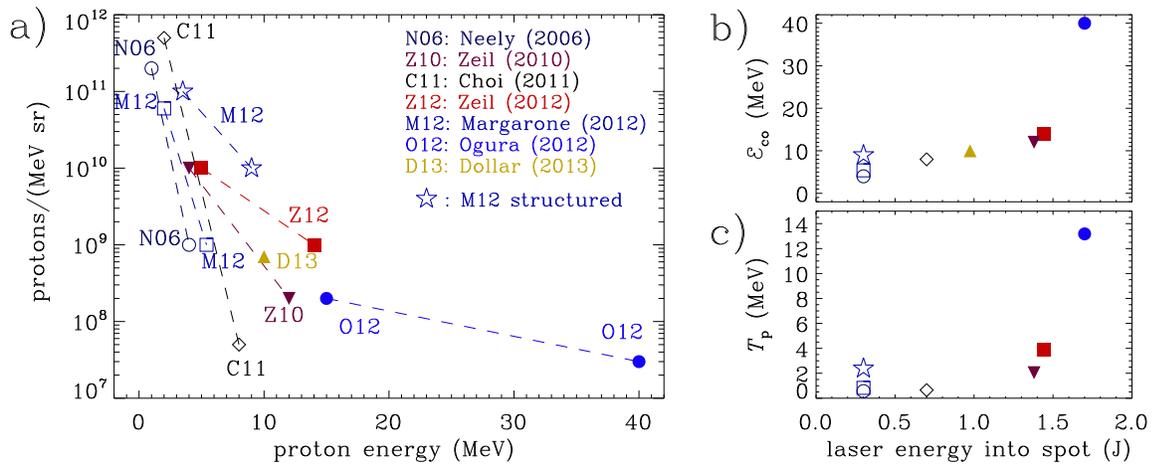}
\caption{Proton energy from experiments using high contrast, sub-100~fs,
sub-10~J laser pulses and thin solid targets. Data are from 
Refs.\cite{neelyAPL06} (N06, empty circles), 
\cite{zeilNJP10} (Z10, filled down-triangles), 
\cite{choiAPL11} (C11, diamonds),
\cite{zeilNC12} (Z12, filled squares), 
\cite{oguraOL12} (O12, filled circles),
\cite{margaronePRL12} (M12, empty squares and stars)
and \cite{dollarPoP13} (D13, filled up-triangles). 
All target are planar foils of various thickness (in the 
$0.05-4.0~\mu\mbox{m}$ range) except for data marked by stars which 
are for a structured target (thin foil covered by sub-micron size spheres).
Data represented by empty and filled symbols are for intensities in the  
$I=(1-5)\times 10^{19}~\mbox{W cm}^{-2}$ and
$I=(0.8-2)\times 10^{21}~\mbox{W cm}^{-2}$ ranges, respectively.
The pulse durations are in the $\tau=(25-40)~\mbox{fs}$ range.
Frame a): spectral density of protons $N_p({\cal E})$ as a function of the
proton energy ${\cal E}$. Dashed lines are simple exponential interpolations 
$N_p({\cal E})=N_{p0}\exp(-{\cal E}/T_p)$
 to observed complete spectra where available.
Frame b): cut-off energy ${\cal E}_{\mbox{\tiny co}}$ as a function of the laser 
pulse energy contained in the laser spot (defined as the area with diameter 
corresponding to the FWHM of the intensity profile).
Frame c): the width parameter $T_p$, from exponential interpolations of the
spectra in frame~a), also as a function of the energy in the spot. 
}
\label{fig:sub10J}
\end{figure}

Since in experiments the observed cut-off energy of protons
${E}_{\mbox{\tiny co}}$ may be determined by the detection threshold, 
a proper comparison of different results may require to compare the full spectra
or, at least, to take into account
both ${E}_{\mbox{\tiny co}}$ and the corresponding number density
$n_{\mbox{\tiny co}}=n_p({E}_{\mbox{\tiny co}})$ with $n_p$ the number of 
protons per energy and solid angle.
Figure~\ref{fig:sub10J}~a) summarizes available information 
by plotting $n_p$ as a function of the proton
energy ${E}$ for the selected experiments.
All reported spectra are well approximated by exponential 
functions $n_p\sim\exp(-E/T_p)$, ``truncated'' at the observed cut-off
${E}_{\mbox{\tiny co}}$.

In Fig.\ref{fig:sub10J}~a) an increasing trend in both ${\cal E}_{\mbox{co}}$
and $T_p$ with the pulse energy is apparent.
In detail, $T_p$ has similar values for the low intensity data (empty symbols)
while for the high intensity data (filled symbols) $T_p$ increases with the
pulse energy and is almost proportional to ${E}_{\mbox{\tiny co}}$.  
It is also  noticeable that the data obtained for higher values of 
pulse intensity and energy correspond to lower values of the number of protons.

Fig.\ref{fig:sub10J}~b) and c) show ${E}_{\mbox{\tiny co}}$ and $T_p$ 
as a function of the pulse energy in the laser spot, defined as the region
corresponding to full-width-half-maximum (FWHM) of the intensity profile. 
This choice is made on the basis of available information on laser spot
size and intensity distribution in 
Refs.\cite{zeilNJP10,zeilNC12,neelyAPL06,choiAPL11,margaronePRL12,oguraOL12,dollarPoP13} 
and allows accounting for different focusing optics and 
plasma mirror efficiencies as well as facilitating the comparison with theoretical models. The amount of energy in the laser spot is 
typically in the 20\%-50\% range of the total laser energy, and the 
values reported in Fig.\ref{fig:sub10J}~b) and c) fall in the $<2$~J 
range.\footnote{Following our choice some quoted values of the ``energy on target'' have been corrected for the comparison. For example, in Refs.\cite{zeilNJP10,zeilNC12} reporting experiments with the DRACO laser, it is reported that 80\% of the laser energy is contained into a spot delimited approximately by the distance at which the intensity falls by $\mbox{e}^{-2}$. Assuming a Gaussian distribution, the energy in the smaller spot defined by the FWHM has been calculated and used in Fig.\ref{fig:sub10J}~b) and c).}

Fig.\ref{fig:sub10J}~b) and c) show that ${E}_{\mbox{\tiny co}}$
is roughly proportional to $T_p$ with an increasing trend of both quantities
with the laser pulse energy. The scaling with the laser intensity $I$ appears
instead to be much weaker than that with energy. The dependence on the 
target thickness does not show a clear trend, with the spread of observed 
values of ${E}_{\mbox{\tiny co}}$ being much smaller than 
range of thicknesses ($0.05-4.0~\mu\mbox{m}$). This latter observation 
is in qualitative agreement with parametric studies which show a relatively
weak dependence on the target thickness in this regime 
\cite{zeilNJP10,zeilNC12,dollarPoP13,prasadAPL11}.

\subsection{TNSA modelling}

The above discussed experimental results may stimulate further advances in
TNSA modelling.
Since the first experiments there has been a demand for models 
yielding reasonably simple scaling laws, able to fit the observed dependencies
on laser and target parameters. 
Ideally such theoretical effort should also reduce the need of additional 
empirical or ``phenomenological'' fitting parameters (such as, e.g., the 
maximum acceleration time) thus raising the predictive power of the models.
Noticeably, several recent works have addressed the
common problem of going beyond the assumption of a simple 
Maxwell-Boltzmann distribution for electrons, because of known problems with, 
e.g., the divergence of the accelerating potential at infinity 
(see e.g. Ref.\cite{macchiRMP13}, Sec.IIIC).

An extension of the static TNSA modelling for arbitrary
electron distribution has been obtained by Schmitz \cite{schmitzPoP12}.
Intriguingly, the use of an ``universal'' electron distribution proposed by 
Sherlock \cite{sherlockPoP09} on the basis of 1D simulations
yields an upper limit of 66~MeV for the
cut-off energy in the case of ultrashort pulse, when the 
ponderomotive scaling is used for the fast electron temperature, i.e.
$T_f=T_{\mbox{\tiny pond}}$ where
\begin{equation}
T_{\mbox{\tiny pond}}=m_ec^2\left[(1+a_0^2/2)^{1/2}-1\right] \; ,
\label{eq:ponderomotive}
\end{equation}
and $a_0=0.85(I\lambda^2/10^{18}~\mbox{W cm}^{-2}\mu\mbox{m}^2)^{1/2}$
is the dimensionless laser amplitude.
A very close upper limit of 65~MeV has been also observed in 
3D PIC simulations for simple flat targets and pulses with energy below 2~J
\cite{dhumieresPoP13}. Thus, going beyond such energy limit could require
the exploitation of particular mechanisms of fast electron generation,
providing a more favourable scaling than (\ref{eq:ponderomotive}).
The use of specially structured targets, to be discussed in the next section, 
is an effort in such direction.

Amongst static TNSA models, 
Passoni et al \cite{passoniPoP13} have recently improved a previous model 
\cite{passoniNJP10}
by accounting for electron recirculation effects in thin targets. 
In this way a dependence of the cut-off value of the accelerating potential 
upon both target thickness and laser energy, which was previously inferred on 
an empirical basis, has been refined and motivated. A preliminary investigation 
(which will be reported elsewhere) with this model of some of the experimental 
data reported in Fig.\ref{fig:sub10J} 
seems to confirm its predicting capabilities, 
highlighting at the same time novel open issues characterizing the TNSA regime 
when ultrashort and tightly focused laser pulses are used.

In the context of dynamic TNSA models, Kiefer et al \cite{kieferPRE13}
have revisited the problem of adiabatic plasma expansion for a steplike 
electron distribution. Deviations due to non-Maxwellian distributions are
shown to be important and relevant to the interpretation of experiments.
Kinetic effects in thin foil expansion have been also recently considered
by Diaw and Mora \cite{diawPRE12}.

For ultrashort pulses the assumption of quasi-equilibrium for electrons
may become questionable.
Zeil et al. \cite{zeilNC12} have experimentally 
investigated the ``pre-thermal'' stage of TNSA 
where the fast electrons in the sheath are out 
of equilibrium: the demonstration of 
proton beam steering by tilting the laser pulse wavefronts 
(resulting in a ``target non-normal'' acceleration) showed that 
the ultrashort laser pulse promptly affects the sheath parameters. 
Non-thermal acceleration of protons in the backward direction (towards the 
laser pulse) has been also investigated with extremely short (5~fs) pulses
\cite{veltchevaPRL12}.
Here we sketch a very simple model of TNSA out of equilibrium.
Assuming that electrons get to the rear surface 
with a momentum $p_0=m_e\gamma_0v_0$ 
as short bunches separated in time by a laser period
$T_L=2\pi/\omega_L$, 
they turn back under the action of a uniform field that reverses
the electron momentum to $-p_0$ and thus gives a kick up to $2p_0$ to
protons at the surface. Since the number of kicks is $\sim (\tau/T_L)$ 
with $\tau$ the pulse duration, the total momentum gain is
$p_i=2p_0(\tau/T_L)$ yielding an energy $p_i^2/2m_i=(2p_0^2/m_i)(\tau/T_L)^2$.
Taking $p_0 \simeq m_ec a_0$  
we obtain a scaling
${\cal E}_p \simeq 2m_ec^2a_0^2(\tau/T_L)^2(m_e/m_p)$
similarly to Ref.\cite{veltchevaPRL12}.
If we rather model the fast electrons as a long bunch of density $n_f$,
the field at the surface will initially grow in time as $E_s=4\pi e n_f v_0 t$.
As far as electron trajectories do not self-intersect, 
an electron emitted at the time $t_0$ will return to the target at
$t_r=t_0+2p_0/E_s(t_0)=t_0+2\gamma_0/(\omega_{pf}^2t_0)$ and the
space-charge field will be largely cancelled at 
$t_s=(2\gamma_0)^{1/2}/\omega_{pf}$ such that $dt_r/dt_0=0$. Protons will thus
gain a momentum $p_i \simeq e\int_0^{t_s}E_s(t)dt=m_ev_0\gamma_0$
and an energy 
${\cal E}_p \simeq m_ev_0^2\gamma_0^2(m_e/2m_p) \sim m_ec^2a_0^2(m_e/m_p)$.
The second model should be more appropriate if $T_L<t_r$, i.e. for 
$\omega_{pf}/\omega_L<(2\gamma_0)^{1/2}/2\pi$ which
means that the fast electron density $n_f$ 
should not largely exceed the critical density $n_c=m_e\omega_L^2/(4\pi e^2)$. 
We thus see that despite the fast scaling with $a_0^2$ the proton energy gain
is limited by the $m_e/m_p$ factor.

For reasons of calculation feasibility almost all the analytical models of
TNSA are one-dimensional, so that multi-dimensional effects may be addressed
only by simulations, mostly performed using PIC codes. The increase of 
supercomputing power now allows large scale simulations to be performed up
to three spatial dimensions (3D). This is particularly relevant for reliable 
quantitative predictions on TNSA because 3D simulations typically show proton
energies lower by a factor of $\gtrsim 2$ with respect to 2D simulations with
the same parameters \cite{dhumieresPoP13,sgattoniPRE12}. At the same time,
TNSA simulations remain very demanding because of issues such as 
large density variations in the sheath, steep gradients at the laser-plasma
interface, and long duration of the acceleration stage. 

\subsection{TNSA in engineered targets}

High-contrast pulses also allow to study the interaction with targets
having nano- and micro-structured surfaces, since the structures may be
not blown out by the prepulse. Thus, target engineering of various type has
been investigated in order to optimize fast electron parameters (such as number,
energy, and divergence) for more efficient TNSA.

Still for sub-10~J pulses,
Margarone et al \cite{margaronePRL12} have investigated  
targets covered by a monolayer of sub-micron spheres. 
These data are also shown in Fig.\ref{fig:sub10J} and the comparison shows
an increase both in ${\cal E}_{\mbox{\tiny co}}$ and in $T_p$ with respect to flat 
targets at small angles of incidence. 
However, measurements with the same type of targets at the SLIC
facility (CEA Saclay) have found TNSA enhancement to be almost absent at large
angles of incidence \cite{floquetXXX13}. 

\begin{figure}[t!]
\includegraphics[width=0.98\textwidth]{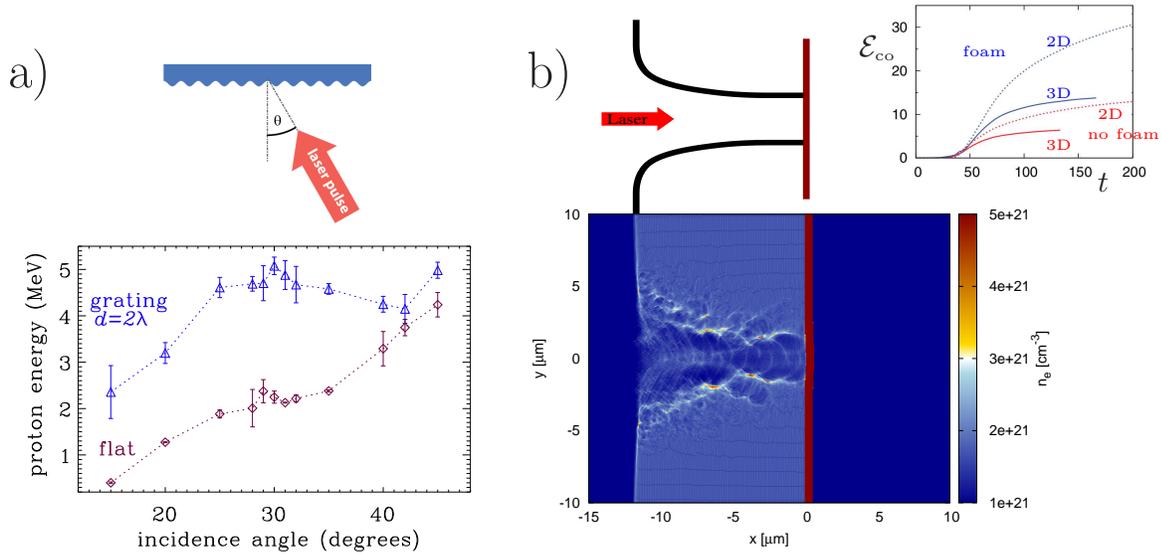}
\caption{Examples of ion acceleration in ``engineered'' targets.
Frame~a): enhancement of cut-off proton energy in laser interaction with grating targets. The cut-off energy increases by a factor 
of $\sim 2.5$ with respect to plane targets around the expected angle for 
resonant excitation of surface waves. Details can be found in 
Ref.\cite{ceccottiXXX13}.
Frame~b): PIC simulation of channel boring in a 
low-density foam layer on the surface of a thin target. The cartoon
shows the schematic of a cone target resembling the channel shape. The plot
in the upper right frame reports the cut-off energy (in MeV) as a function of 
time (in fs) for 2D and 3D simulations, with and without the foam layer.  
The foam layer density and thickness are 
$n_e=2n_c=3.4 \times 10^{21}~\mbox{W cm}^{-2}$ and $d=2\lambda=1.6~\mu\mbox{m}$.
The laser pulse has peak amplitude $a_0=10$ ($I=10^{20}~\mbox{W cm}^{-2}$), 
25~fs duration and $3~\mu\mbox{m}$ waist radius . 
}
\label{fig:gratings}
\end{figure}

At SLIC an increase in ${\cal E}_{\mbox{\tiny co}}$ has been also observed in 
``grating'' targets
\cite{sgattoniEPS,ceccottiXXX13} having a periodical surface modulation
with sub-wavelength depth. For this experiment, Fig.\ref{fig:gratings}~a) shows
${\cal E}_{\mbox{\tiny co}}$ as a function of the incidence angle for both grating 
and plane
targets having a depth of $\sim 20~\mu\mbox{m}$. The laser pulse energy in 
the FWHM spot and the intensity were 0.3~J and 
$2.5 \times 10^{19}~\mbox{W cm}^{-2}$, respectively.
The maximum enhancement is observed for the angle at which the resonant
excitation of surface waves is expected. The achieved values of 
${\cal E}_{\mbox{\tiny co}}$
are somewhat limited by the large value of the target thickness, since the 
surface structuring of very thin foils is difficult. At the same time, the
details of proton emission provide a very useful diagnostic for the laser
absorption and electron acceleration mechanisms in the presence of 
microstructures.\footnote{Using a particular form of cryogenic structured ``snow'' targets  Zigler et al \cite{ziglerPRL13} have observed up to 21~MeV protons with a 2~J laser pulse at MBI Berlin. Both the number and emittance of such energetic protons appear to be lower than what is observed in TNSA experiments, and the acceleration is effective for low pulse contrast. Hence the acceleration mechanism is presumably quite different from TNSA.}

Use of ``shaped'' targets (having a structure size of the order of the
laser spot, i.e. a few $\mu\mbox{m}$)
and high laser contrast allow also to
achieve higher proton energies for pulse of relatively long duration 
($0.1-1$~ps) and high energy ($10^2-10^3$~J).  
Using specially designed microcone targets and the 80~J, 700~fs, high contrast
pulse of the TRIDENT laser at LANL, Gaillard et al \cite{gaillardPoP11}
were able to detect about $10^7$ protons/MeV at a cut-off energy 
${E}_{\mbox{\tiny co}}=67.5~\mbox{MeV}$, to be compared with
${E}_{\mbox{\tiny co}} \simeq 50~\mbox{MeV}$ in flat targets.  
The cut-off energy was higher than the long-lasting record of 58~MeV
observed with $>400$~J pulses at the LLNL petawatt by 
Snavely et al \cite{snavelyPRL00}. However, it has to be noticed that
in the spectrum  measured in LLNL experiments ${\cal E}_{\mbox{\tiny co}}$ 
corresponded to about $10^9$ protons/MeV, a number for which 
the spectrum measured at TRIDENT for microcone targets yields about 40~MeV. 
Thus, the higher energy cut-off observed 
might be partly attributed to an increased sensitivity in the detector. 
Nevertheless, the cone target results also compare favourably with respect to 
those of Robson et al \cite{robsonNP07} which found protons up to 
55~MeV at similar proton/MeV numbers using 400~J, $1-8$~ps pulses from the 
VULCAN petawatt laser at RAL, and similar value of the target thickness. 

The performance of the cone targets appears to be based on efficient 
electron acceleration in cone geometry \cite{klugeNJP12}. Indeed, some 
features of the acceleration mechanism such as coupling with the cone walls
at local grazing incidence and existence of an optimal cone length are
also apparent in simulations of channel drilling in a near-critical density
plasma \cite{sgattoniPRE12}. In a thin target covered with a near-critical 
layer, efficient electron acceleration may yield electron temperatures up
to $T_f\simeq 3T_{\mbox{\tiny pond}}$ and up to a three-fold increase in 
${E}_{\mbox{\tiny co}}$ with respect to targets without foam layers: 
see the simulation results shown in Fig.\ref{fig:gratings}~b) where the 
effect of 3D geometry versus 2D is also shown. 
Simulations of this scheme have been performed so
far for ultrashort, low-energy pulses because of computational limitations.
Also experimentally the scheme has been tested in the short-pulse regime and
with foam thicknesses suitable to imply a gain at low intensity conditions
\cite{zaniC13,prencipeEPS}.
Further investigations with longer pulses are needed to compare with cone 
targets. 

\subsection{Proton focusing and spectral manipulation}

Suitably shaped targets have been also used since the first experiments to
focus the proton beam, without the need of an external steering system and
allowing the focusing of large number of protons as it would be required 
by fast ignition and high energy density physics applications.
Recently, progress on this side has been obtained by Bartal et al 
\cite{bartalNP12} on TRIDENT using cone targets similar to those designed for 
fast ignition research. Focusing by concave targets has been investigated
also at LULI \cite{chenPRL12} revealing an issue of proton beam filamentation
and emittance degradation. 

Another possibility of present interest is based on the use of the strong 
quasistatic magnetic fields generated during the laser-plasma interaction
and related to the process of sheath formation and evolution itself. 
Early analysis of experiments on ultrathin foils traced back spectral 
modulations in the low-energy part of the proton spectrum to the effect
of magnetic fields $>$10~MG generated at the rear surface within tens of
fs \cite{robinsonNJP09}. Measurements based on the proton probing
technique \cite{sarriPRL12} gave both evidence of the generation of fields
of several tens of MG  by the ``fountain effect'' of fast electrons
\cite{macchiXXX12} and of their capability to either focus or defocus 
multi-MeV protons. Very recent experiments on the TITAN laser at LLNL have
tested this concept \cite{fuchsEEAC13}.

The engineering of target geometry and composition has been also investigated
with the aim to manipulate the TNSA spectrum and obtain suitably monoenergetic
distributions reducing the need of external filtering devices
(see e.g. sec.III.E.2 in Ref.\cite{macchiRMP13}). 
However, recently there seems to have been little progress on this side.
For instance a seemingly promising 
approach based on targets with nanolayers deposited on their rear side
\cite{hegelichN06,schwoererN06} has not been further developed. 
These issues, as well as the uncertainties in reaching via TNSA-based 
schemes the high energies $>100$~MeV/nucleon required by some applications, 
motivate the search for alternate mechanisms which we describe in the next
sections.

\section{Towards Radiation Pressure Acceleration}

\subsection{RPA in thick targets: hole boring} 

The radiation pressure of an intense laser pulse pushes the plasma surface
inwards, creating a bow-shaped deformation. For thick targets, such that 
the deformation front does not reach the rear side of the target 
during the pulse, this regime has been commonly named as ``hole boring'' (HB). 
Assuming a uniform density, cold plasma target
with mass density $\rho=Am_pn_i=(A/Z)m_pn_e$, the fastest ions
have an energy per nucleon 
(see sec.IV.A.1 of Ref.\cite{macchiRMP13} and references therein)
\begin{equation}
{\cal E}_{\mbox{\tiny HB}}=2m_pc^2\frac{\Pi}{1+2\Pi^{1/2}} \; ,
\qquad
\Pi=\frac{I}{\rho c^3}=\frac{Z}{A}\frac{m_e}{m_p}\frac{n_c}{n_e}a_0^2 \; .
\end{equation}
corresponding to twice the velocity of the receding surface 
$v_{\mbox{\tiny HB}}=c\Pi^{1/2}/(1+\Pi^{1/2})$.
Recently a theory of HB-RPA including effects of temperature, pulse envelope and
density inhomogeneity has been presented by Levy et al \cite{levyHEDP13}
who also discuss possible relevance of HB experiments to astrophysical 
scenarios, such as Poynting flux-dominated outflows.

Simulations of HB-RPA show that the fast ions may have a narrow energy spread 
and thus lead to a non-thermal, peaked ion energy spectrum particularly
for a circularly polarized pulse at normal incidence since fast electron 
generation and, consequently, TNSA are strongly suppressed. 
Protons with energy $\simeq 1~\mbox{MeV}$ and $\simeq 10\%$ energy spread
were observed in an experiment at Brookhaven National Laboratory (US)
\cite{palmerPRL11}
where a circularly polarized CO$_2$ pulse of amplitude $a_0=0.5$ and 
a gas H jet target were used allowing values of $n_e/n_c \gtrsim 1$.
The potential advantages of this experimental scheme are the reduction of 
the electron density to increase ${\cal E}_{\mbox{\tiny HB}}$, the 
pure proton plasma and the suitability for high repetition rate.

The scaling with density makes the transition to HB-RPA hard to observe in 
solid targets where $n_e\gtrsim 10^2n_c$. However, either the presence of a 
significant preplasma or the use of advanced targets at low density 
(liquid hydrogen jets would be particularly suitable)
may allow to work with $n_e/n_c \gtrsim 1$ also with optical lasers. 
The approach might be of interest especially for next-generation short-pulse
lasers aiming at very large values of $a_0$ with modest pulse energy 
and possibly allowing high repetition rate operation.
Related simulation studies and suggestions for experiments have been
reported in Refs.\cite{macchiNIMA10,robinsonPoP11,robinsonPPCF12}.

\subsection{RPA in thin targets: light sail}

For thin targets of thickness $\ell\ll v_{HB}\tau$, RPA enters into the 
``light sail'' (LS) mode in which the whole target is accelerated. For LS, the 
scaling is now with the pulse fluence $I\tau$ (total energy per unit surface)
divided by the areal density of the target 
(see sec.IV.A.2 of Ref.\cite{macchiRMP13} and references therein):
\begin{equation}
{\cal E}_{\mbox{\tiny LS}}
=m_pc^2\frac{(\Omega\tau)^2}{2(\Omega\tau+1)} \; ,\qquad
\Omega=\frac{2I}{\rho\ell c^2}
      =T_L^{-1}\frac{Z}{A}\frac{m_e}{m_p}\frac{a_0^2}{\zeta} \; , \qquad
\zeta=\pi\frac{n_e}{n_c}\frac{\ell}{\lambda} \; .
\label{eq:LS}
\end{equation}
The LS operation is limited by the onset or relativistic transparency when
$a_0>\zeta$, thus the ``optimal'' matching of the areal density with the pulse 
amplitude occurs for $a_0=\zeta$.\footnote{Strictly speaking this condition is derived on the basis of a simple model of delta-like foil plasmas and has been verified by simulations using circularly polarized, ultrashort pulses (see e.g. Ref.\cite{macchiPRL09}). In more general cases the ``optimal'' thickness value determined by transparency effects may have a more complex dependence on the laser pulse parameters (see e.g. Refs.\cite{macchiNJP10,dongPRE03,esirkepovPRL06}).} 
This condition is presently accessible thanks
to manufacturing techniques for foils as thin as a few nm, for which typically
$\zeta \lesssim 10$. Of course ultrahigh contrast is necessary for this type
of experiment. Posing $a_0=\zeta$ and $Z/A=1/2$ in Eq.(\ref{eq:LS}) and 
assuming non-relativistic ions ($\Omega\tau\ll 1$), we may write a simplified
scaling law for LS-RPA using optimal thickness targets: 
\begin{equation}
{\cal E}^{\mbox{\tiny (opt)}}_{\mbox{\tiny LS}}
=\frac{m_e^2c^2}{8m_p}a_0^2\left(\frac{\tau}{T_L}\right)^2
\simeq 2.3~\mbox{MeV} \times \frac{I\tau^2}{10^{-6}~\mbox{J s cm}^{-2}}
= 2.3~\mbox{MeV} I_{20}\tau^2_{100} \; ,
\label{eq:LSopt}
\end{equation}
where $I_{20}$ and $\tau_{100}$ are the intensity and pulse duration in 
units of $10^{20}~\mbox{W cm}^{-2}$ and $100~\mbox{fs}$, respectively.
The more favourable scaling with pulse duration than with intensity is due to 
the latter being limited by the transparency threshold.  
Eq.(\ref{eq:LSopt}) shows that energies $>100$~MeV per nucleon are within 
reach of present-day petawatt systems. 

Several recent experiments have investigated the interaction with ultrathin 
foils obtaining indications of the onset of LS-RPA. 
Using the VULCAN petawatt pulse 
($\sim 800~\mbox{fs}$, $I=0.5-3 \times 10^{20}~\mbox{W cm}^{-2}$)
Kar et al observed spectral peaks for both protons
and heavier (e.g. carbon) ions, 
with energies up to 10~MeV/nucleon,
and inferred from data a scaling of the peak 
energy as $(\Omega\tau)^2$ in accordance with Eq.(\ref{eq:LS}) for 
$\Omega\tau\ll 1$. 
Similar results have been obtained by Aurand et al
 \cite{aurandNJP13} using the much shorter pulse 
(27~fs, $I=0.2-6 \times 10^{19}~\mbox{W cm}^{-2}$) of the JETI laser
at Jena. Common to both experiments was a weak dependence upon the laser
polarization and a separation in the spectrum between proton and carbon
peaks. Thus, the acceleration occurs in the presence of fast electrons, and
the peak separation might be due to additional TNSA-like effect giving an extra
boost preferentially to the lighter protons. Similar spectra were also obtained
by Steinke et al \cite{steinkePRSTAB13} at MBI Berlin
using circular polarization only.
This suggests that side effects such as the tight focusing of the laser
pulse ($f/2.5$ for Ref.\cite{steinkePRSTAB13}) cause significant heating
also for circular polarization, because of both the non-negligible 
longitudinal electric field components and the target deformation which leads
to locally oblique incidence. For even tighter focusing ($f/1$), using the
HERCULES laser at $I=2\times 10^{21}~\mbox{W cm}^{-2}$
Dollar et al \cite{dollarPRL12} found that finite spot size effects 
almost cancel out the differences between linear and circular polarization
and are unfavourable for RPA. 

Recent results by Kim et al \cite{kimXXX13} obtained with the 30~fs PULSER
laser at GIST using ultrathin targets, linear polarization and 
$f/4$ focusing with an intensity range 
$I=0.5-3.3 \times 10^{20}~\mbox{W cm}^{-2}$
show a more complex scenario. Non-thermal, peaked spectra are observed for
C ions while the proton spectrum is broad, modulated and extends up to
a cut-off energy ${\cal E}_{\mbox{\tiny co}}=45~\mbox{MeV}$ 
for the thinnest target (10~nm). The number of accelerated protons appears to
be higher by an order of magnitude than those observed for thicker targets
by Ogura et al \cite{oguraOL12} at a similar cut-off energy. 
In addition, varying the intensity showed a
transition from a ${\cal E}_{\mbox{\tiny co}}\propto I^{1/2}$ to 
${\cal E}_{\mbox{\tiny co}}\propto I$ at an intermediate intensity (with the
peak energy of carbon spectra showing a similar trend). 
This trend (also apparent in the peak energy of C) is 
similar to that observed by Esirkepov et al in multi-parametric 
simulations \cite{esirkepovPRL06} and attributed to the progressively dominant
role of RPA. For the thinnest target values, however, one would also expect 
a strong effect of relativistic transparency, which might account for the 
broad proton spectra. Comparison with 
PIC simulations also suggests that the modulations in the energy spectrum are 
due to a transverse instability of the foil target, similarly to previous
measurements on VULCAN \cite{palmerPRL12}. 
 
As an overall comment on the above mentioned experiments, the importance of
RPA effects at the highest intensities (or in more generally suitable 
conditions) is apparent and the evidence of the ``fast'' quadratic scaling
with fluence is promising. Still, a long way towards optimization remains. 
In particular, the non-thermal spectra observed for thin foils are relatively 
broad, in contrast to the simple picture of LS-RPA as a thin plasma bunch
moving at a coherent velocity, which would correspond to a perfectly 
monoenergetic spectrum. The need for wide laser spots to prevent 
target bending and excessive heating would imply large amounts of laser
energy. 
Many observed features need to be fully understood,
especially the RPA dynamics with multiple species. Such effects could be
also exploited to stabilize the acceleration, and more in general target
engineering might also be pursued to tailor the spectrum 
(see several references in Ref.\cite{macchiRMP13}, sec.IV.A.2).

\subsection{RPA as a route towards relativistic ions}

In LS-RPA the conversion efficiency 
$\eta=2\beta_{\mbox{\tiny LS}}/(1+\beta_{\mbox{\tiny LS}})$ 
reaches unity when the sail velocity $\beta_{\mbox{\tiny LS}}\rightarrow 1$.
However, the scaling with the pulse fluence is less favourable in this regime.
Assuming again $a_0 \simeq \zeta$ at the optimal condition we obtain
${\cal E}^{\mbox{\tiny (opt)}}_{\mbox{\tiny LS}}
=(m_e c^2 a_0/2)\left({\tau}/{T_L}\right)$.
Based on this estimate
the laser fluence required to reach 
${\cal E}_{\mbox{\tiny LS}}=m_pc^2$ is $1.5 \times 10^8~\mbox{J cm}^{−2}$.
The corresponding energy would depend strongly on the possible requirement of 
a sufficiently wide spot to minimize effects of target deformation: 
as an example, taking a $10~\mu\mbox{m}$ diameter would require 120~J energy.
The slow rate of energy gain in the laboratory frame as 
${\cal E}_{\mbox{\tiny LS}} \sim (\Omega t)^{1/3}$ also represents a possible issue
since the acceleration distance might be large enough that pulse diffraction
must be taken into account.  

\begin{figure}
\includegraphics[width=0.99\textwidth]{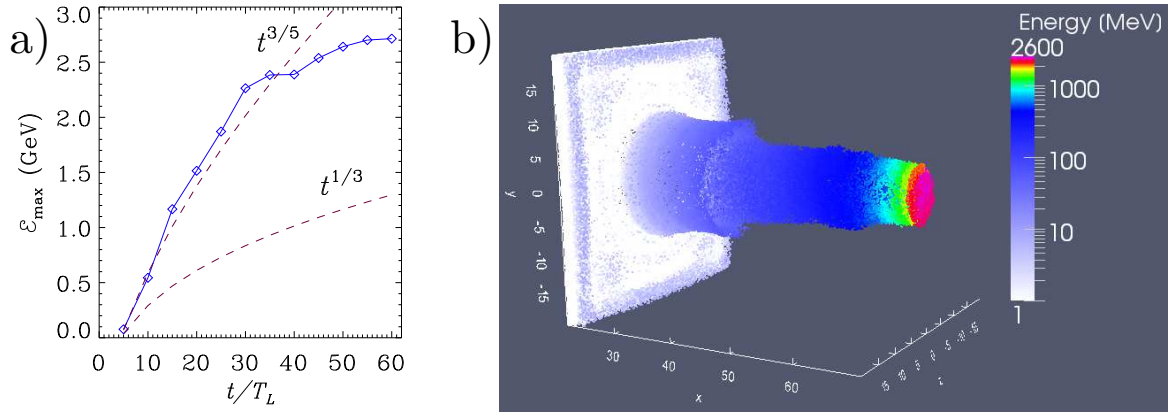}
\caption{3D simulations of LS-RPA. 
Frame a): maximum ion energy versus time (in laser cycles).
Frame b):  distribution in space and energy of ions 
directed into a cone with $10^{\circ}$ full aperture. 
The target has an initial thickness
$\ell=\lambda$ and density $n_e=64n_c$ 
($=1.1 \times 10^{23}~\mbox{cm}^{-3}$ for $\lambda=0.8~\mu\mbox{m}$)
corresponding to $\zeta \simeq 200$. 
The laser pulse has $9T_L$ (24~fs) FWHM duration (``$\sin^2$'' envelope), 
$a_0=198$ peak amplitude ($I=8.5 \times 10^{22}~\mbox{W cm}^{-2}$), 
$6\lambda$ ($4.8~\mu\mbox{m}$) waist radius, and circular polarization.}
\label{fig:LS-RPA}
\end{figure}

While these issues are certainly challenging, theory and simulation have 
identified possible regimes of high gain LS-RPA in the relativistic ion regime.
Using 2D simulations and analytical modelling 
Bulanov et al \cite{bulanovPRL10} have shown how the target rarefaction might
dynamically reduce the areal density on the axis. This allows a 
faster energy gain ${\cal E}_{\mbox{\tiny LS}} \sim (\Omega t)^{3/5}$ at the
expense of a lower number of accelerated ions, provided that a transition 
to transparency (partly prevented by the decrease of laser pulse frequency
in the target frame) does not occur. 
Tamburini et al \cite{tamburiniPRE12} have further investigated this regime
with fully 3D simulations, finding that
the energy maximum observed in 3D simulations is \emph{higher}
than in 2D, since the effects of rarefaction are stronger.
Exploiting the computing power of the FERMI supercomputer at CINECA
(Bologna, Italy) the simulations of Ref.\cite{tamburiniPRE12} have been
extended to longer times. Fig.\ref{fig:LS-RPA}~a) shows the maximum
energy as a function of time, which is well fitted by the $(\Omega t)^{3/5}$
scaling up to a time of $t=30T_L$. At later times the growth
becomes slower, presumably because of the onset of strong transmission of the
pulse through the target. The energy of $\simeq 2.6$~GeV at $t=60T_L$ is 
more than six times the prediction of the 1D formula (\ref{eq:LS})  for 
${\cal E}_{\mbox{\tiny LS}}$.
These results show that multi.GeV energies might be reached with pulse
energies of the order of 1~kJ, 
combined with pulse durations and focusing suitable to 
approach the $10^{23}~\mbox{cm}^{-3}$ intensity level 
(in the case of Fig.\ref{fig:LS-RPA} a waist radius of $6\lambda$ and
a pulse duration of 24~fs were used). 
The spatial distribution in Fig.\ref{fig:LS-RPA}~b) shows 
that the most energetic ions are collimated into a narrow cone.

\section{Acceleration in the relativistic transparency regime}

While the onset of relativistic transparency in a thin target stops the
LS-RPA stage, it opens up a regime of volumetric interaction with the
near-critical density plasma which may lead to high ion energy. In particular,
proton cut-off energies exceeding 100~MeV have been communicated 
\cite{hegelichBAPS11}. Several experiments in the transparent regime 
have been performed with TRIDENT at LANL (see e.g. Ref.\cite{hegelichNF11} 
for a review of results oriented to fast ignition ICF research). 
Most of the observation are explained in the context of the 
Break-Out Afterburner (BOA) model which is actually a relatively complex 
mechanism, involving strong instability stages which lead to strong heating of 
electrons and coupling to ions. While scalings with laser and target parameters
have probably to be fully characterized yet, BOA has shown promising results
for what concerns cut-off energies and number of ions. A recent work 
\cite{jungNJP13} shows in particular an efficient bulk acceleration of C ions,
producing up to $5 \times 10^{11}$ particles with an high efficiency of
6\%. 

A typical signature of the onset of the transparent regime 
is that the highest ion numbers and energies are not detected along the laser 
axis at some angle (e.g. $10^{\circ}$ for Ref.\cite{jungNJP13}). 
This effect is also observed in 3D simulations analogous to that shown in
Fig.\ref{fig:LS-RPA} but for linear polarization. In such a case the 
transmission of the laser pulse is stronger than for circular polarization and
the maximum energy is lower.

\section{Collisionless Shock Acceleration}

An experiment with a similar set-up to Ref.\cite{palmerPRL11}, i.e. 
using a CO$_2$ laser and a hydrogen jet target was also performed
with the NEPTUNE laser at UCLA (US) \cite{haberbergerNP12}. 
In this case the laser polarization
was linear and the intensity of was four times higher than in 
Ref.\cite{palmerPRL11}. Very narrow peaks in the proton spectrum
were observed up to 22~MeV with $\simeq 10^7$ protons~MeV$^{-1}$sr$^{-1}$
in the peak, in striking contrast with 
$>10^{12}$ protons~MeV$^{-1}$sr$^{-1}$ at $\simeq 1$~MeV for 
Ref.\cite{palmerPRL11} which gives evidence of a different acceleration
mechanism.

The UCLA experiment was interpreted on the basis on the 
Collisionless Shock Acceleration (CSA) model. The fast heating of electrons
and density steepening by the laser pulse drive collisionless shock
waves at the velocity $v_s=Mc_s$ (where the sound velocity 
$c_s=(ZT_e/Am_p)^{1/2}$ for a non-relativistic plasma and electrons in Boltzmann
equilibrium, and $M>1$ is the Mach number). These shocks 
are associated to ions reflected by the shock front. 
As far as $v_s$ is constant the reflected ions form a monoenergetic beam 
at the velocity $2v_s$. If the shock is sustained by fast electrons
($T_e=T_f$) high values of $v_s$ and of ion energies 
$m_p(2v_s)^2/2=2m_pv_s^2$ may be reached. Much theoretical and simulation work
has been performed to unfold the details of the acceleration process and
in particular to tailor the plasma density profile obtaining a suitably smooth
plasma gradient, with the additional aim
to suppress TNSA in favour of CSA 
(see \cite{fiuzaPoP13} and references therein).

Scaling CSA beyond 100~MeV/nucleon will require either the development of
CO$_2$ lasers with higher energy or a suitable set-up with optical lasers,
using target at densities around $n_c \simeq 10^{21}~\mbox{cm}^{-3}$ which is
higher than what is commonly achieved with gas jets.
Recent experiments were performed at LULI using exploding foil plasmas
(pre-heated by long pulses) which turn underdense at the time of the 
interaction with a short intense pulse \cite{dhumieresJPCS10}.
With this method, broad proton
spectra up to a cut-off energy of $7.1$~MeV with $\simeq 10^{9}$ 
protons~MeV$^{-1}$sr$^{-1}$ have been obtained with a 8~J, 400~fs, 
$5 \times 10^{18}~\mbox{W cm}^{-2}$ laser pulse
\cite{gauthierEPS12}.

\begin{figure}
\includegraphics[width=0.98\textwidth]{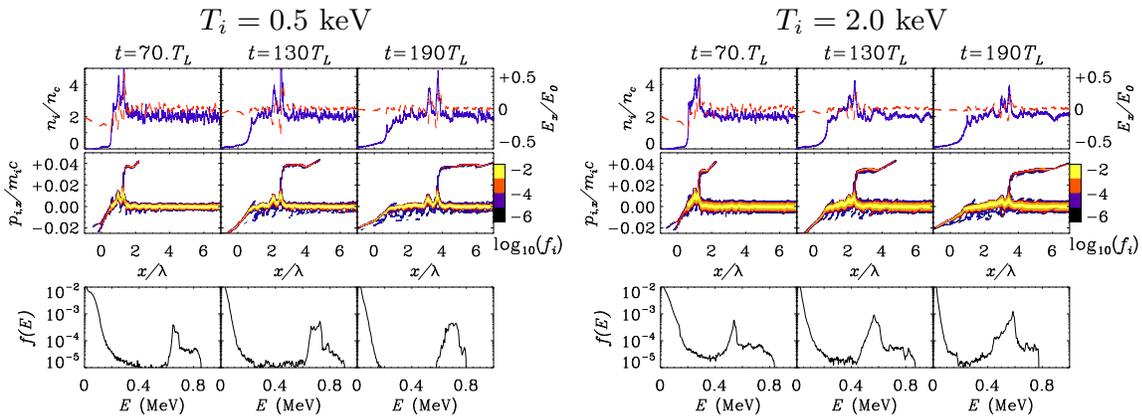}
\caption{One-dimensional PIC simulation of shock acceleration. 
Results are shown for two runs with identical parameters but the initial ion 
temperature $T_i$ (0.5 and 2~keV, respectively). 
For both simulations snaphshots at three different times (in units of the 
laser period $T_L$) are shown.
The laser pulse impinges from the left and reaches at $t=0$
the plasma boundary at $x=0$.
The top row shows the profiles of ion density $n_i$ and electric field $E_x$. 
The middle row shows the $f_i(x,p_x)$ projection of the ion phase space
distribution. 
The bottom row shows the ion energy spectrum.
For higher $T_i$ the shock velocity decreases and the spectrum
progressively broadens towards lower energies.
The laser and plasma parameters are $a_0=1$, $\tau=55T_L$, $n_e=2n_c$.
The density and the electric field are normalized to $n_c$ and $m_e\omega c/e$,
respectively.
}
\label{fig:shock}
\end{figure}

A possibly limiting factor in CSA is that high efficiency may be
not compatible with keeping a monoenergetic spectrum. This is because
the energy gained by the accelerated ions is at expense of the shock wave
energy, so that the shock velocity decreases. In turn, the velocity of the
reflected ions decreases as well, causing a progressive chirp of the spectrum 
towards low energies  \cite{macchiPRE12}.
To illustrate the basics of the effect and its dependence on 
the initial velocity distribution of the background ions, 
we show in Fig.\ref{fig:shock} the results of two
1D PIC simulations with identical parameters but different initial 
temperature $T_i$. For higher $T_i$, a larger number of ions is reflected 
from the shock, causing a stronger loading of the shock front which 
velocity decreases. 
As a consequence the peak  in the spectrum broadens in time.
This effect may explain why the monoenergetic
spectra of Ref.\cite{haberbergerNP12} correspond to quite low number of 
protons. This is also an issue for PIC simulations because a very low
energy tail needs to be resolved, requiring very high numbers of computational
particles.

\section{Conclusions}

There has been considerable recent progress in laser-driven ion acceleration. 
In the last two years the observed energy cut-offs have increased with
short pulse, low energy laser systems reaching several tens on MeV, close
to or even exceeding what was obtained with large petawatt systems. 
Progress have been also obtained on the side of conversion efficiency, 
total number of particles, and energy spread. It might be argued that 
on each side the most promising results have been obtained in different 
experiments with no acceleration mechanism at present seeming able to combine 
all desirable qualities. However, there appears to be room for further 
optimization, with each mechanism being ultimately most suitable for a specific 
application.

Ultrahigh pulse contrast has played a crucial role in experiments with
solid targets, allowing for instance the use of ultrathin foils, and probably
has allowed more controlled conditions and more reproducible 
observations. Promising results have been also achieved by the use of 
structured targets, with ion acceleration also providing an insight on the
peculiar interaction physics. The development of innovative targetry has
a crucial importance also for the perspective of high-repetition rate operation.

The further development of high-power laser systems will allow soon to extend
the present investigations to unexplored regimes. Relevant laser 
development may not only involve optical systems with femtosecond 
pulses but also CO$_2$ infrared lasers which are of potential interest for 
radiation pressure and shock acceleration in gas targets. 

Significant advances have also been made on the side of modelling of 
experiments and understanding of relevant physics. Still, however, several
open questions remain and in several cases theoretical scalings with laser
and target parameters are still uncertain. Further progress on this side
is highly desirable, especially in order to provide reliable 
predictions on experiments with next-generation laser systems. 
In particular there is a strong need, common to different acceleration 
mechanisms, for massively parallel simulations able to perform more realistic
investigations, in a three-dimensional geometry and over large scales of
space and time.

\ack
A.M. thanks J.~Fuchs (LULI, \'Ecole Polytechnique, France), 
K.~Zeil (HZDR, Germany), 
E.~d'Humi\`eres (CELIA, France), 
D.~Margarone (ELI-Beamlines, Czech Republich) 
and Chul Min Kim (GIST, Korea) for providing 
additional information and comments on published and unpublished results.
The grating targets data of Fig.\ref{fig:gratings} have been obtained in
an experiment at the SLIC facility (CEA Saclay, France) 
funded from LASERLAB-EUROPE 
(grant agreement no. 284464, EU's 7FP, proposal n.SLIC001693) 
and were analysed by T.~Ceccotti and V.~Floquet.
The contribution of P.~Londrillo to ALaDyn code development and simulations is
gratefully acknowledged.
Access to FERMI supercomputer at CINECA was awarded by PRACE via the project 
``LSAIL''.  
Support from EPSRC, grant EP/E035728/1 (LIBRA consortium)
and from the Italian Ministry of University and Research via
the FIR project ``SULDIS'' is acknowledged.

\end{document}